\documentclass[aps,pre,twocolumn,superscriptaddress]{revtex4}

\usepackage{amsmath}
\usepackage{amssymb}
\usepackage{amsbsy}
\usepackage{makeidx}
\usepackage{epsf}
\usepackage{graphicx}
\usepackage{amsfonts}
\usepackage{subfigure}
\usepackage{epstopdf}
\usepackage{mathrsfs}
\usepackage{xcolor}

\newcommand{\revision}[1]{\textcolor{black}{#1}}

\begin{document}
	\title{\revision{Topological edge solitons and their stability in a nonlinear Su-Schrieffer-Heeger model}}
	\author{Y.-P.\ Ma}
	\email{yiping.m@gmail.com}
	\affiliation{Department of Mathematics, Physics and Electrical Engineering,
		Northumbria University, Newcastle Upon Tyne, NE1 8ST, United Kingdom}
	\author{H.\ Susanto}
	\email{hadi.susanto@ku.ac.ae}
	\affiliation{Department of Mathematics, Khalifa University of Science and Technology, PO Box 127788, Abu Dhabi, United Arab Emirates}
	\affiliation{Department of Mathematical Sciences, University of Essex, Wivenhoe Park, Colchester CO4 3SQ, United Kingdom}
	

	\begin{abstract}
		We study continuations of topological edge states in the Su-Schrieffer-Heeger model with on-site cubic (Kerr) nonlinearity\textcolor{black}{, which is a 1D nonlinear photonic topological insulator (TI)}. Based on the topology of the underlying spatial dynamical system, we establish the existence of nonlinear edge states (edge solitons) for all positive energies in the topological band gap. We discover that these edge solitons are stable at any energy when the ratio between the weak and strong couplings is below a critical value. Above the critical coupling ratio, there are energy intervals where the edge solitons experience an oscillatory instability. \textcolor{black}{Though our work focuses on a photonic system, we also discuss the broader relevance of our methods and results to 1D nonlinear mechanical TIs.}
	\end{abstract}
	
	\maketitle
	
	\section{Introduction}
	
	\begin{figure}[tbh]
		\centering
		\includegraphics[width=7cm]{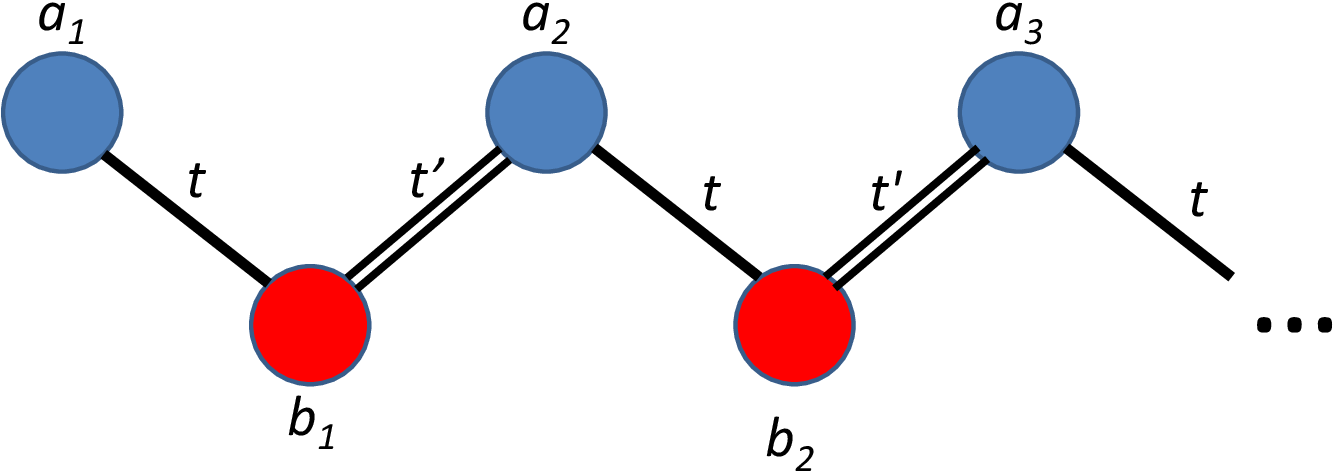}
		\caption{Schematic representation of the semi-infinite one-dimensional lattice in the SSH model.}
		\label{fig1} 
	\end{figure}
	
	Topological insulators (TIs) are exotic materials that behave like an insulator in the bulk and a conductor on the edge~\cite{Moore-2010,Hasan-Kane-2010,Qi-Zhang-2010}. Prototypical TIs exhibit nontrivial topology in their bulk wave functions, which guarantees the existence of robust edge states in the bulk band gap(s)~\cite{CTSR16-RMP}. Though TIs were originally proposed in 2D and subsequently generalized to 3D, the simplest TI already exists in 1D as the Su-Schrieffer-Heeger (SSH) model~\cite{ssh79}. Here we consider an orbital version of the SSH model on a 1D lattice with two sites per unit cell and different intra- ($t$) and inter-cell ($t'$) hopping amplitudes~\cite{stje17}; see Fig.~\ref{fig1}. This model is described by the following Hamiltonian with chiral symmetry 
	\begin{equation}
	\label{Hamiltonian}
	\hat{H}=\sum_{j=1}^\infty t\hat{a}_{j}\hat{b}^{\dagger}_{j} + t'\hat{a}^{\dagger}_{j+1}\hat{b}_{j} + h.c.,
	\end{equation}
	where $\hat{a}^{\dagger}_{j}$ ($\hat{b}^{\dagger}_{j}$) is the creation operator on the site $a_{j}$ ($b_{j}$). \textcolor{black}{A standard calculation shows that the bulk wave functions exhibit a phase $\Phi$ that is a function of the momentum $k$, and the bulk spectrum consists of two bands $\pm(|t-t'|,t+t')$ separated by a gap. The {\it winding number} $\mathcal{W}$ of the phase $\Phi\left(k\right)$ over the Brillouin zone can be shown to be a topological invariant that is trivial ($\mathcal{W} = 0$) for $t>t'$ and non-trivial ($\mathcal{W} = 1$) for $t<t'$~\cite{Delplace2011}.} The principle of bulk-edge correspondence then guarantees that in the latter case only, there exists a topologically protected edge state in the semi-infinite chain in Fig.~\ref{fig1} that terminates at $a_1$. This edge state exists in the bulk band gap with zero energy, and has an exponentially decaying envelope $|a_j|\sim\Lambda^{j}$ and $|b_j|=0$, where $\Lambda=t/t'$. \textcolor{black}{Note that the chiral symmetry of the Hamiltonian in Eq.~\eqref{Hamiltonian} is essential for the existence of the winding number ${\cal W}$ as a topological invariant~\cite{CTSR16-RMP}.}
	
	The principles underlying quantum TIs have been fruitfully adapted to photonic (electromagnetic)~\cite{Ozawa19_RMP,SmLe19_arXiv} and phononic (mechanical)~\cite{Huber16_NP,Liu20_AFM} systems. The governing differential equations of these systems often exhibit nonlinearity, which can interact with topology to create new dynamical effects. \revision{A prominent feature of such nonlinear TIs is localized states that bifurcate from the topologically protected edge state with zero-energy in the linear limit. Hereafter these are referred to as topological edge solitons and will be the focus of this paper. Though photonic TIs only gained popularity in recent years, there are earlier studies on 1D waveguide arrays with alternating spacings~\cite{ViJo09_PRA,Kanshu12_OL}. This setup is essentially identical to the SSH model, but the topological properties of such systems remain largely unexplored in these earlier studies. In~\cite{ViJo09_PRA}, bulk gap solitons are found in the SSH model with on-site cubic (Kerr) nonlinearity, and their stability is analyzed systematically. In~\cite{Kanshu12_OL}, both bulk and surface (edge) gap solitons are found in the SSH model with on-site saturable nonlinearity, but there are no explicit discussions of topological edge solitons.}
	
	\revision{More recently,} there have been several proposals to include inter-site or on-site nonlinearity in the SSH model, with the latter considered particularly relevant to photonics~\cite{SmLe19_arXiv}. In~\cite{Hadad16,Hadad17}, self-induced topological transitions are achieved in SSH models whose inter-cell coupling depends on the local dimer energy. In~\cite{soln17}, topological gap solitons with nontrivial chirality are found in the SSH model with on-site cubic nonlinearity. This model reduces to the nonlinear Dirac model in the continuum limit, where topological edge states and gap solitons are found to share a common origin~\cite{smir19}. In the presence of chiral symmetry breaking, this model is found to exhibit rich topological physics that entail new edge and bulk solitons~\cite{tulo20}. Topological gap solitons are also found in a mechanical SSH model with on-site cubic nonlinearity~\cite{Chau21_PRB} and a photonic SSH model with on-site saturable nonlinearity~\cite{Guo20_OL}. Other nonlinearities also find prominent applications including topological lasers~\cite{Malz18_NJP,Malz18_OE} and topological circuits~\cite{Enge17_PRL,Kotw19_arXiv} among others.
	
	The SSH model with on-site Kerr nonlinearity can be regarded as the prototypical model for realizing nonlinear topological localized states. Here we explain the origin of topological gap solitons in this model, including both bulk and edge solitons, based on the phase space geometry of the underlying discrete spatial dynamical system \textcolor{black}{(DS)}. We use numerical continuation to obtain the complete set of topological edge solitons, and compute their stability to reveal the inherent limit of energy storage on the edge. We also present analytical approximations in the small coupling ratio limit and the continuum limit. Our aim is to characterize edge solitons systematically, since only particular instances of this solution set have been reported to our knowledge~\cite{smir19,tulo20}. Our method can be generalized to many other nonlinear TIs to expand their tunability through the discovery of new solutions.
	
	\section{Spatial dynamics}
	
	The governing equation of the SSH model \eqref{Hamiltonian} with Kerr nonlinearity is
	\begin{align}
	\begin{split}
	&i\dot{a}_j=tb_j+t'b_{j-1}+|a_j|^2a_j,\\ 
	&i\dot{b}_j=ta_j+t'a_{j+1}+|b_j|^2b_j,
	\end{split}
	\label{gov}
	\end{align}
	$j\in\mathbb{Z}^+$, $b_0=0$, and $\dot{\square}$ denotes derivative in the propagation distance (time) $z$. By scaling, we have taken the nonlinear coefficient to be 1 and will take $t'=1$ in our numerical computations below. Eq.~\eqref{gov} conserves the power $P=\sum_{j} |a_j|^2+|b_j|^2$ and the Hamiltonian
	\begin{equation*}
	H=\sum_{j}t(a_j^*b_j+b_j^*a_j)+t'(a_j^*b_{j-1}+b_j^*a_{j+1})+\frac{1}{2}(|a_j|^4+|b_j|^4).
	\end{equation*}
	Hereafter we assume $t<t'$ such that the linear limit of Eq.~\eqref{gov} is topologically nontrivial.
	
	We look for standing waves of Eq.~\eqref{gov} in the form of $a_j=A_j\exp(-iEz)$, $b_j=B_j\exp(-iEz)$, where $E$ is the propagation constant (energy). Assuming that $A_j$ and $B_j$ are both real without loss of generality~\cite{syaf12}, Eq.~\eqref{gov} reduces to the following 2D invertible map in $j$:
	\begin{equation}\label{eq:rev-map}
	    EA_j=tB_j+t'B_{j-1}+A_j^3,\, EB_j=tA_j+t'A_{j+1}+B_j^3.
	\end{equation}
	Due to Hermiticity, Eq.~\eqref{eq:rev-map} is reversible under the transformation $R:(j,A,B)\rightarrow(-j,B,A)$, which implies that its phase portrait in the $(A,B)$-plane is symmetric with respect to the section $\mathrm{fix}(R):=\{(A,B)|A=B\}$~\cite{RoQu92,LaRo98}. In the band gap $|E|<t'-t$, the only equilibrium on the $\mathrm{fix}(R)$ section is the origin $O:(0,0)$ with eigenvalues
	\begin{equation}\label{eq:lambda}
	\lambda_1=-\tan(\theta/2),\quad \lambda_2=-\cot(\theta/2),
	\end{equation}
	and the corresponding eigenvectors
	\begin{equation}
	    v_1=(-\cot(\phi/2),1),\quad v_2=(-\tan(\phi/2),1),
	\end{equation}
	where
	\begin{equation*}
	\theta=\arcsin\left(\frac{2t't}{t'^2+t^2-E^2}\right),\, \phi=\arcsin\left(\frac{2tE}{t'^2-t^2-E^2}\right).
	\end{equation*}
	Since $\theta\in(0,\pi/2)$, the equilibrium $O$ is a saddle whose stable and unstable eigenspaces are respectively defined by $(\lambda_1,v_1)$ and $(\lambda_2,v_2)$. The polar angle $-\phi/2$ of $v_1$ lies in $(-\pi/4,0)$ for $E\in(0,t'-t)$ and $(0,\pi/4)$ for $E\in(t-t',0)$.
	
	\begin{figure}[tbh]
	    \centering
	    \begin{tabular}{cc}
	       \includegraphics[width=8cm]{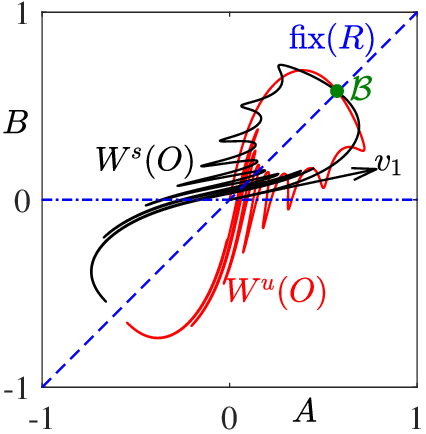}  \\ \includegraphics[width=8cm]{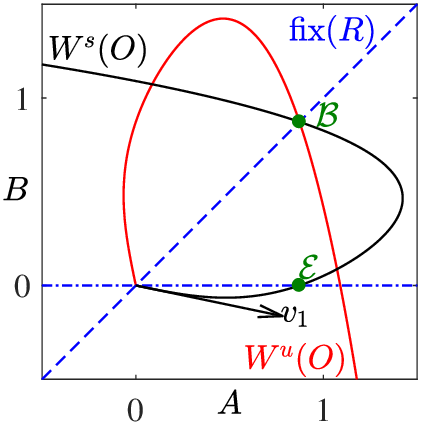}
	    \end{tabular}
	    \caption{Phase portrait of the 2D map given by Eq.~\eqref{eq:rev-map} at $t=0.6$ and (top panel) $E=-0.2$; (bottom panel) $E=0.2$. The solid curves are the stable manifold $W^s(O)$ and the unstable manifold $W^u(O)$, the dashed line is the symmetry section $\mathrm{fix}(R)$, the dot-dashed line is the boundary section $B=0$, and the arrow is the stable eigenvector $v_1$. The points ${\cal B}$ and ${\cal E}$ respectively represent a bulk soliton and an edge soliton.}
	    \label{fig:SSH_man}
	\end{figure}
	
	The origins of both bulk and edge solitons can be explained in the $(A,B)$ phase plane. In Fig.~\ref{fig:SSH_man}, we show the stable manifold $W^s$ and the unstable manifold $W^u$ of the saddle $O$ at $E<0$ (top panel) and $E>0$ (bottom panel). A bulk soliton is a reversible homoclinic orbit consisting of the backward and forward iterates of a transverse intersection between $W^s(O)$ and $\mathrm{fix}(R)$~\cite{C98_PhysD}, while an edge soliton consists of the forward iterates of a transverse intersection between $W^s(O)$ and the section $B=0$ that represents the boundary condition. For both $E<0$ and $E>0$, $W^s(O)$ emanates from $O$ along the $v_1$ direction and bends towards $\mathrm{fix}(R)$ to form a bulk soliton represented by ${\cal B}$. For $E>0$ only, $B=0$ is sandwiched between $v_1$ and $\mathrm{fix}(R)$, so $W^s(O)$ crosses $B=0$ to form an edge soliton represented by ${\cal E}$. This sandwiching property provides a simple topological criterion for the existence of edge solitons in nonlinear TIs in general.
	
	We note that for $E>0$, although the bulk soliton ${\cal B}$ and the edge soliton ${\cal E}$ belong to the same invariant manifold $W^s(O)$, their spatial profiles generally do not coincide except in the continuum limit~\cite{smir19}. We also note that for $E<0$, $W^s(O)$ wiggles after crossing $\mathrm{fix}(R)$ at ${\cal B}$ and intersects $\mathrm{fix}(R)$ and $B=0$ to form additional bulk and edge solitons, some of which have been discovered computationally in~\cite{tulo20}. Finally, we note that Eq.~\eqref{gov} with $t'=t$ is simply the focusing discrete nonlinear Schr\"odinger (NLS) equation~\cite{Kevr09_Book}. Therefore, our analysis shows that ``topologisation'' of classical models in nonlinear waves can greatly enrich their solution sets.
	
	\begin{figure}[tbh]
		\centering
		{\includegraphics[width=9cm]{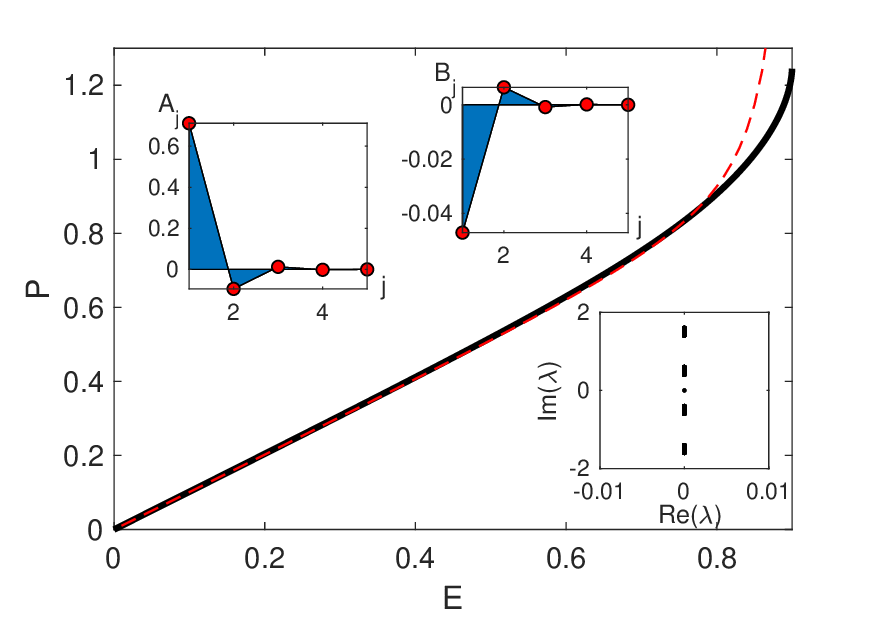}}
		\caption{Power $P$ vs.\ energy $E$ of edge solitons with $t=0.1$. The solid and dashed curves are from numerics and approximation \eqref{pa}, respectively. The insets show the solution profile and its spectrum for $E=0.5$ (lines are from numerics, and circles are from approximation \eqref{appr}).}
		\label{fig2} 
	\end{figure}
	
	\section{Numerical continuation}
	
	Hereafter we focus on the primary family of edge solitons ${\cal E}$ that bifurcate towards $E>0$. We can use numerical continuation to obtain these solutions, which essentially tracks ${\cal E}$ as $E$ varies without explicitly computing $W^s(O)$. Once a solution is found, its linear stability is computed through solving the corresponding eigenvalue problem obtained from substituting $a_j=(A_j+\epsilon\hat{a}_j\exp(\lambda z))\exp(-iEz)$, $b_j=(B_j+\epsilon\hat{b}_j\exp(\lambda z))\exp(-iEz)$ and linearizing about $\epsilon=0$. In Fig.\ \ref{fig2}, we present the power $P$ for varying $E$ with $t=0.1$. An example profile and its spectrum in the complex plane are also shown. \textcolor{black}{As $E$ increases from 0 to $t'-t$, the stable spatial eigenvalue $\lambda_1$ in Eq.~\eqref{eq:lambda} decreases from $-t/t'$ to $-1$. This implies that the edge soliton decays more slowly in space as $E$ increases and becomes a bulk state as $E$ tends to $t'-t$, where the solution branch terminates as shown in Fig.~\ref{fig2}.} \revision{Here we note the different conventions that we have used for defining the spectrum: the spectrum of an edge soliton consists of growth rates $\lambda$, while the spectrum of the Hamiltonian $\hat{H}$ in Eq.~\eqref{Hamiltonian} consists of angular frequencies; these are consistent with the standard conventions in the nonlinear waves and quantum physics literature. Thus, in the limit $E\rightarrow0$, the spectrum of the edge soliton reduces to $i$ times the spectrum of the Hamiltonian $\hat{H}$ in Eq.~\eqref{Hamiltonian}.}

	\begin{figure}[tbh]
		\centering
		{\includegraphics[width=9cm]{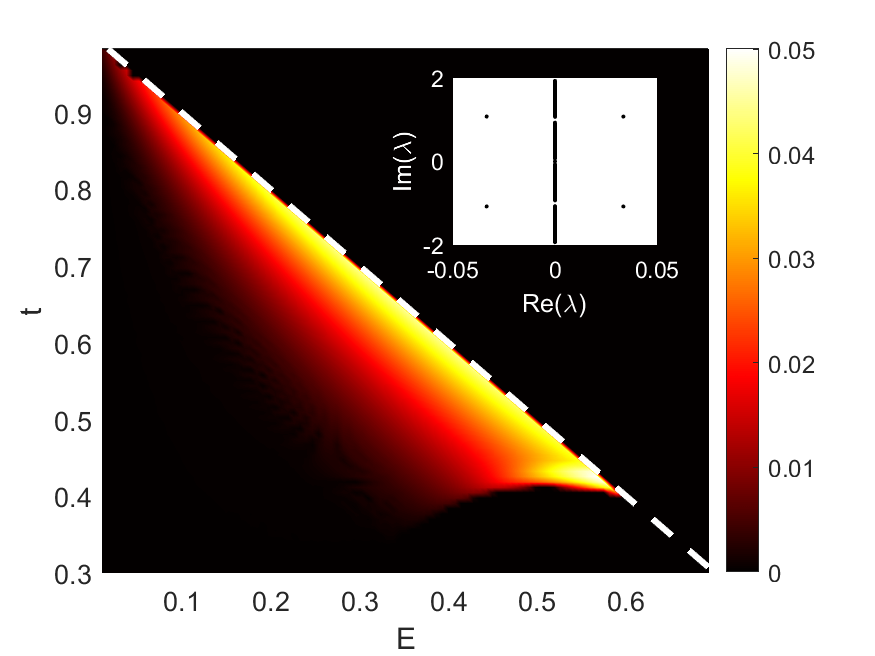}} 
		\caption{\textcolor{black}{The maximum of the real part of the spectrum of edge solitons in the $(t,E)$ plane.} Region with nonzero spectrum corresponds to unstable solutions. Edge solitons only exist below the dashed white line $E=t'-t$. The inset shows 
		the spectrum of an edge soliton at $E=0.5$ and $t=0.43$, i.e., it suffers oscillatory instability.}
		\label{fig3} 
	\end{figure}
	
	We also study the effect of varying the coupling $t$ and the energy $E$ in general. For every $t\in(0,t')$ and $E\in(0,t'-t)$, there exists a unique edge soliton. We show in Fig.\ \ref{fig3} the critical spectrum of 
	these edge solitons in the $(t,E)$ plane\textcolor{black}{, defined as the maximum of the real part of the spectrum}. The edge soliton is stable for all $E$ when $t<t_c\approx0.35$, but becomes unstable over certain interval(s) of $E$ when $t>t_c$. The nature of the instability is oscillatory as seen in the inset of Fig.~\ref{fig3}. The critical eigenvalue appears following the collision of two eigenvalues that bifurcate from the lower edge of the uppermost band and the upper edge of the next-uppermost one. This is 
	typical of the oscillatory instability appearing in gap solitons \cite{bara00,derk05}. As $t$ increases, we obtain more and more unstable eigenvalues appearing.
	
	\begin{figure}[tbh]
		\centering
		\begin{tabular}{cc}
		\includegraphics[width=9cm]{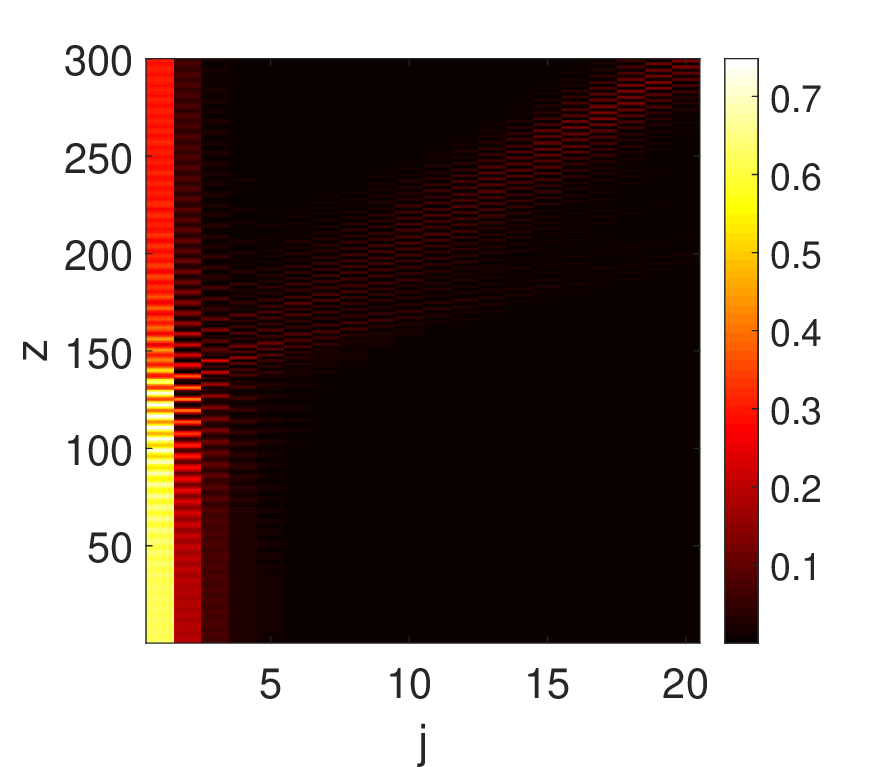} \\ \includegraphics[width=9cm]{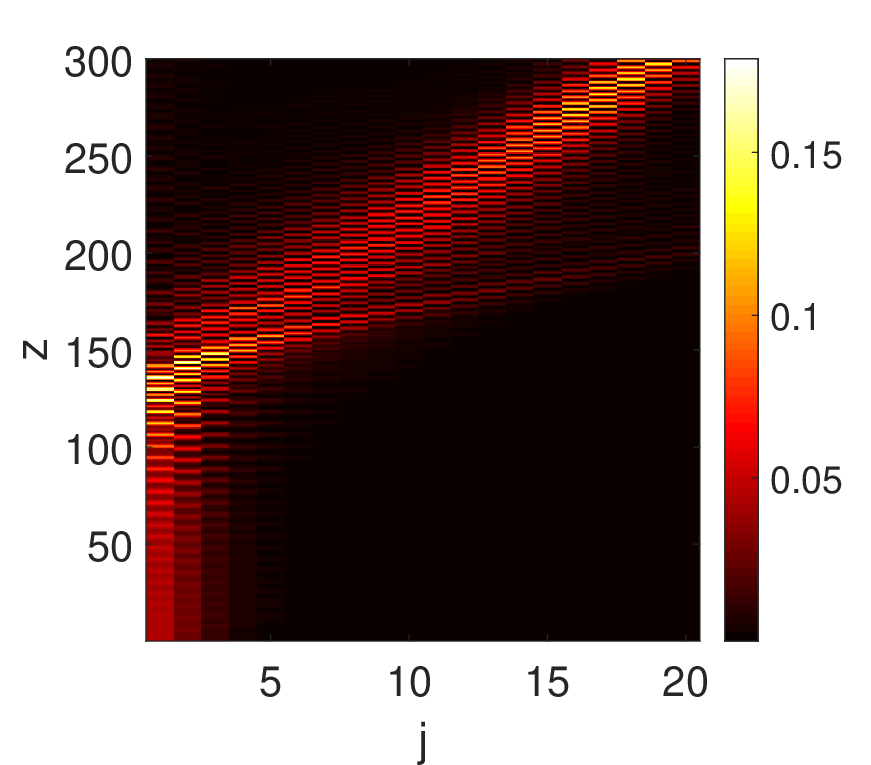} 
		\end{tabular}
		\caption{
		Time dynamics of an edge soliton corresponding to the inset of Fig.\ \ref{fig3}, i.e., $t=0.43$. Shown is the intensity $|a_j|^2$ (top panel) and $|b_j|^2$ (bottom panel). The oscillatory nature of the instability is eminent, where radiation in the form of a localized packet travelling towards the bulk can be clearly seen.}
		\label{fig4} 
	\end{figure}
	
	We simulate the typical dynamics of the oscillatory instability. Using an exact edge soliton as the initial condition with random perturbation that is proportional to the field amplitude, we show the time evolution of Eq.~\eqref{gov} in Fig.\ \ref{fig4}. \textcolor{black}{At $t=0.43$, the initial edge soliton with $E=0.5$, which is inside the unstable region in Fig.~\ref{fig3}, gradually loses energy by releasing a localized wave packet into the bulk, and eventually evolves into a stable edge soliton with smaller power, 
	which is just outside the unstable region in Fig.~\ref{fig3}. This release process approximately in $100<z<150$ features temporal oscillations with an angular frequency of about 1.08, consistent with the imaginary part of the unstable eigenvalues in the inset of Fig.~\ref{fig3}. This process} is almost the exact opposite of the mechanism discussed in \cite{smir19,MaRe19_arXiv}, where a bulk soliton is launched towards the edge to increase the energy of an edge soliton. Our result implies that such proposals to couple the bulk and edge modes are robust only when the edge soliton is stable. We also note that a similar instability is observed using a different realization of the Kerr nonlinearity in coupled fiber loops that implement a discrete 1D quantum walk~\cite{BiWi19_PRA}.

	\section{Analytical approximations}
	
	For the coupling ratio $|\Lambda|\ll1$, we can approximate the edge state by 
	\begin{equation}
	A_j\approx \tilde{A}(-L)^{j-1},\quad B_j\approx \tilde{B}(-L)^j,
	\label{appr}
	\end{equation}
	with, using the assumption that only the equation for $a_1$ is nonlinear and keeping the remaining sites linear, 
	\begin{align*}
	&L=\frac{t^2+t'^2-E^2-\sqrt{E^4-2E^2(t^2+t'^2)+(t^2-t'^2)^2}}{2tt'},\\
	&\tilde{A}=\sqrt E, \quad \tilde{B}=(t'-t/L)/\sqrt E.
	\end{align*}
	In the limit $E\to0$, we can calculate that $L\to\Lambda=t/t'$. Using \eqref{appr}, the power is given by 
	\begin{equation}
	P\approx (\tilde{A}^2+\tilde{B}^2L^2)/(1-L^2).
	\label{pa}
	\end{equation}
	This compares well with the numerics as seen in Fig.~\ref{fig2}.
	
	The stability in the limit $\Lambda\to0$ can be established using perturbation theory, exploiting dimers of the system; see a similar problem considered in, e.g., \cite{kiri16}. The method, that has been standard by now, can be applied here to show that in that limit the edge soliton is stable. 
	
	On the other hand, near the Dirac point $(t,E)=(t',0)$, the edge soliton becomes slowly varying, i.e., we have the continuum limit~\cite{LeDe16_APX}. In that case, denoting $\square=a,b$, we introduce a small parameter $\epsilon$ and take: $t=t'-E-\epsilon\delta$, $\square_j=(-1)^j\sqrt{\epsilon}\widehat{\square}_j\exp(-iEz)$, $\widehat{\square}_j=\widehat{\square}(j\epsilon)=\widehat{\square}(x)$ such that $\widehat{\square}_{j\pm1}=\widehat{\square}(x\pm\epsilon)=\widehat{\square}(x)\pm\epsilon\widehat{\square}_x+\mathcal{O}(\epsilon^2)$, $E=\epsilon\widehat{E}$, and a slow time scale $\widehat{z}=\epsilon z$. After dropping the hats, Eq.~\eqref{gov} becomes the nonlinear Dirac (NLD) equation~\cite{HaCa09_PhysD}
	\begin{align}
	\begin{split}
	&i\dot{a}+Ea=(-E-\delta)b+t'b_x+|a|^2a,\\
	&i\dot{b}+Eb=(-E-\delta)a-t'a_x+|b|^2b,
	\end{split}
	\label{gov2}
	\end{align}
	with 
	boundary conditions $b(0)=0$ and $\lim_{x\to\infty}a(x)=0$.
	
	\begin{figure}[tbh]
		\centering
		\begin{tabular}{cc}
		\includegraphics[width=9cm]{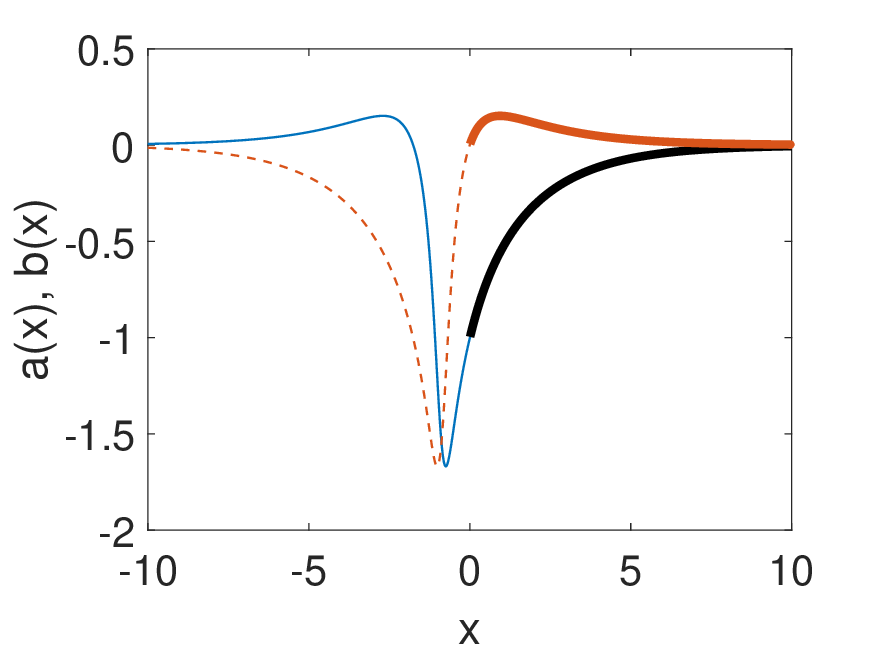} \\ \includegraphics[width=9cm]{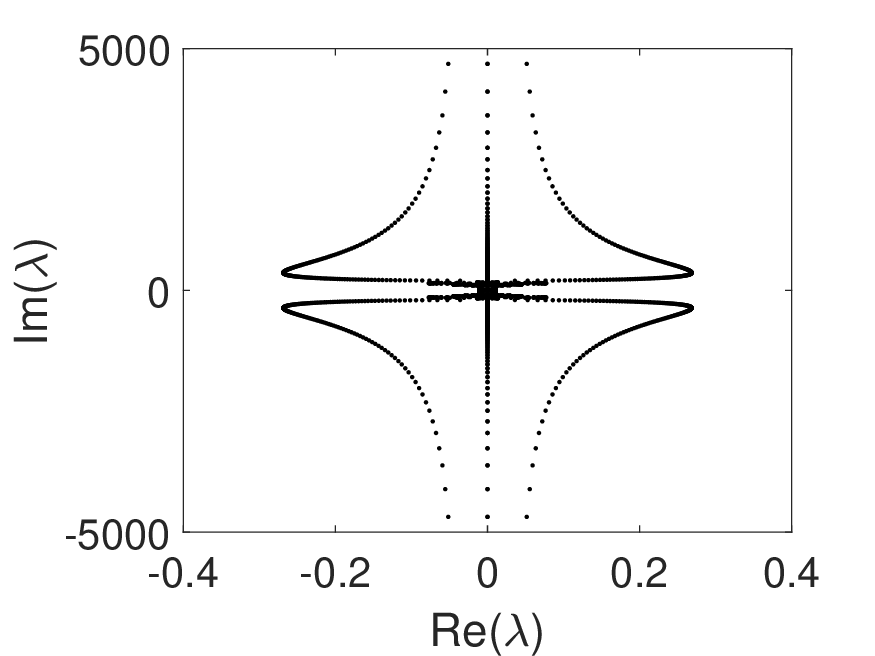}
		\end{tabular}
		\caption{(top panel) Bulk soliton of \eqref{gov2} at $E=0.5$ and $\delta=0.1$. Solid and dashed lines are the fields $a$ and $b$, respectively. The part of the profile that makes an edge soliton is highlighted as thick lines. (bottom panel) Spectrum of the edge soliton in the complex plane, showing  that it is highly unstable, in the sense that there is a continuous unstable spectrum.}
		\label{fig5} 
	\end{figure}
	
	As shown in~\cite{smir19}, Eq.~\eqref{gov2} admits an exact analytical solution on the infinite domain that describes traveling bulk solitons. At zero velocity, the stationary bulk soliton has a spatial orbit that passes through $b=0$, so part of this orbit describes a stationary edge soliton. An example stationary profile 
	is plotted in Fig.\ \ref{fig5}(a); note that the bulk soliton can be identified as a reversible homoclinic orbit invariant under $(x,a,b)\rightarrow(-x,b,a)$~\cite{C98_PhysD}.
	
	The stability of solitons in the NLD equation~\eqref{gov2} is harder to analyze than in the NLS equation due to the absence of the Vakhitov-Kolokolov criterion~\cite{CuKe16_PRL}. This task can be challenging even numerically because simple discretization (finite difference) of the eigenvalue problem can generate a large number of spurious unstable eigenvalues; see, e.g.,   \cite{bara98,bara00} for the stability of gap solitons in a similar system. Delicate methods, such as Chebyshev interpolation method \cite{chug06} and Evans function techniques \cite{derk05}, can be used to capture the actual instability. Here, we have used the Chebyshev method to solve the eigenvalue problem corresponding to Eq.~\eqref{gov2}. We show the spectrum in Fig.\ \ref{fig5}(b), where we obtain that the edge soliton in Fig.~\ref{fig5}(a) experiences an oscillatory instability, which is consistent with the top left corner of Fig.\ \ref{fig3}. 
	
	\textcolor{black}{
	\section{Comparison with mechanical systems}
	In view of recent work on edge/interface solitons in 1D nonlinear mechanical TIs~\cite{Chau21_PRB,Temp21_arXiv}, we compare our 1D nonlinear photonic TI with its mechanical counterpart, namely the mechanical SSH model with on-site cubic nonlinearity~\cite{Chau21_PRB}. The key difference is that a photonic system can be reduced to a spatial DS using a simple ansatz with a single harmonic in time (with angular frequency $E$ in our notation) even for strong nonlinearity, but this is generally impossible in a mechanical system except for weak nonlinearity. In particular, edge solitons in~\cite{Chau21_PRB} (referred to as nonlinear normal modes in~\cite{Temp21_arXiv}) generally consist of infinitely many harmonics in time.
	}
	
	\textcolor{black}{
	Still, the model in~\cite{Chau21_PRB} can be viewed as a spatial DS for the Fourier components in time on each lattice site. After a Galerkin truncation that keeps only the lowest harmonic, the resulting spatial DS is similar to Eq.~\eqref{eq:rev-map} in our model. Without this truncation, the spatial DS in~\cite{Chau21_PRB} is higher-dimensional and therefore exhibits bifurcation diagrams different from Eq.~\eqref{eq:rev-map} in our model. Most notably, the branches of edge solitons in Figs.~2 \& 3 of~\cite{Chau21_PRB} penetrate into the bulk bands, while edge solitons in our model exist strictly within the band gap.
	}
	
	\textcolor{black}{
	Though the concept of spectral stability of nonlinear waves is similar in photonic and mechanical systems, the details of the analyses are different. The stability of an edge soliton in our model depends on the eigenvalues of a time-independent linear operator, while the stability of an edge soliton in~\cite{Chau21_PRB} depends on the Floquet multipliers of a time-periodic linear operator. Interestingly, the stability trend is exactly opposite between our model and the model in~\cite{Chau21_PRB} with stiffening nonlinearity: edge solitons in our model are generally stable {\it below} a critical energy (frequency), while edge solitons in~\cite{Chau21_PRB} are generally stable {\it above} a critical energy (frequency) as shown in Fig.~3 of~\cite{Chau21_PRB}. Thus, a natural question is whether edge solitons can be stabilized for all energies (frequencies) using a hybrid between these two models.
	}
	
	\section{Conclusion and Outlook}
	
	We have studied the existence and stability of topological edge solitons in the SSH model with on-site Kerr nonlinearity. The phase space topology implies that edge solitons exist for all positive energies in the topological band gap. These edge solitons are completely stable below a critical coupling ratio, but can experience an oscillatory instability above it. Our results are pertinent to photonic experiments using, for example, 1D topological arrays of coupled nonlinear resonators~\cite{dobr18}. Our results can also be realized in the 1D Gross-Pitaevskii equation with a periodic dimer potential, which reduces to the nonlinear SSH model in the tight-binding limit~\cite{soln17}.
	
	The topological edge solitons in the nonlinear SSH model correspond directly to extended edge states in 2D nonlinear TIs such as nonlinear photonic graphene~\cite{ACZ13_PRA}. These 2D extended edge states inherit the instability of their 1D counterparts longitudinally, but are subject to an additional transverse instability~\cite{LRPS16_PRA}. In addition to these essentially 1D localized states, we expect that new intrinsically 2D localized states in 2D nonlinear TIs can also be discovered using numerical continuation in the spatial dynamics framework.

	\section*{Acknowledgement}
	
	YM acknowledges support from a Vice Chancellor's Research Fellowship at Northumbria University.


\begin{thebibliography}{99}

\bibitem[{Moore(2010)}]{Moore-2010}
\bibinfo{author}{J.~E. Moore},
\newblock \bibinfo{title}{The birth of topological insulators},
\newblock \bibinfo{journal}{Nature} \bibinfo{volume}{464}
 \bibinfo{pages}{194}  (\bibinfo{year}{2010}).
\bibitem[{Hasan and Kane(2010)}]{Hasan-Kane-2010}
\bibinfo{author}{M.~Z. Hasan and C.~L. Kane},
\newblock \bibinfo{title}{Colloquium: topological insulators},
\newblock \bibinfo{journal}{Reviews of Modern Physics} \bibinfo{volume}{82}
 \bibinfo{pages}{3045}  (\bibinfo{year}{2010}).
\bibitem[{Qi and Zhang(2011)}]{Qi-Zhang-2010}
\bibinfo{author}{X.-L. Qi and S.-C. Zhang},
\newblock \bibinfo{title}{Topological insulators and superconductors},
\newblock \bibinfo{journal}{Reviews of Modern Physics} \bibinfo{volume}{83}
 \bibinfo{pages}{1057}  (\bibinfo{year}{2011}).

\bibitem[{Chiu et~al.(2016)}]{CTSR16-RMP}
\bibinfo{author}{C.-K. Chiu, J.~C. Teo, A.~P. Schnyder, and S.~Ryu},
\newblock \bibinfo{title}{Classification of topological quantum matter with symmetries},
\newblock \bibinfo{journal}{Reviews of Modern Physics} \bibinfo{volume}{88}
 \bibinfo{pages}{035005}  (\bibinfo{year}{2016}).

\bibitem{ssh79} W. P. Su, J. R. Schrieffer, and A. J. Heeger, Solitons in polyacetylene, Phys. Rev. Lett. 42, 1698 (1979).

		\bibitem{stje17} P. St-Jean, V. Goblot, E. Galopin, A. Lemaître, T. Ozawa, L. Le Gratiet, I. Sagnes, J. Bloch, and A. Amo, Lasing in topological edge states of a one-dimensional lattice, 
		Nature Photonics 11, no. 10, 651 (2017).
		\bibitem{Delplace2011} P. Delplace, D. Ullmo, and G. Montambaux, Zak phase and the existence of edge states in graphene, Physical Review B 84, 195452 (2011).

\bibitem{Ozawa19_RMP} T. Ozawa, H.M. Price, A. Amo, N. Goldman, M. Hafezi, L. Lu, M.C. Rechtsman, D. Schuster, J. Simon, O. Zilberberg, and I. Carusotto, Topological photonics, Reviews of Modern Physics, 91(1), p.015006 (2019).
\bibitem{SmLe19_arXiv} D. Smirnova, D. Leykam, Y. Chong, and Y. Kivshar, Nonlinear topological photonics, Applied Physics Reviews, 7(2), 021306 (2020).
\bibitem{Huber16_NP} S. D. Huber, Topological mechanics, Nature Physics, 12(7), 621 (2016).
\bibitem{Liu20_AFM} Y. Liu, X. Chen, and Y. Xu, Topological phononics: From fundamental models to real materials, Advanced Functional Materials, 30(8), 1904784 (2020).

\bibitem{ViJo09_PRA} R. A. Vicencio and M. Johansson, Discrete gap solitons in waveguide arrays with alternating spacings, Physical Review A, 79(6), 065801 (2009).

\bibitem{Kanshu12_OL} A. Kanshu, C. E. R\"uter, D. Kip, V. Shandarov, P. P. Beličev, I. Ilić, and M. Stepić, Observation of discrete gap solitons in one-dimensional waveguide arrays with alternating spacings and saturable defocusing nonlinearity, Optics letters, 37(7), 1253-1255 (2012).

		\bibitem{Hadad16}
		Y. Hadad, A. B. Khanikaev, and A. Alu, Self-induced topological transitions and edge states supported by nonlinear staggered potentials,
		Physical Review B, 93(15), 155112 (2016).
		\bibitem{Hadad17}
		Y. Hadad, V. Vitelli, and A. Alu, Solitons and propagating domain walls in topological resonator arrays,
		ACS Photonics, 4(8), 1974-1979 (2017).
		\bibitem{soln17} D.D. Solnyshkov, O. Bleu, B. Teklu, G. Malpuech, Chirality of topological gap solitons in bosonic dimer chains, Phys. Rev. Lett. 118, 023901 (2017).
		\bibitem{smir19} D. Smirnova, L. Smirnov, D. Leykam, and Y. Kivshar, Topological edge states and gap solitons in the nonlinear Dirac model,
		Laser \& Photonics Reviews, 13(12), 1900223 (2019).
	    \bibitem{tulo20} T. Tuloup, R. W. Bomantara, C. H. Lee, and J. Gong, Nonlinearity induced topological physics in momentum space and real space,
	    Physical Review B, 102(11), 115411 (2020).
	    \bibitem{Chau21_PRB} R. Chaunsali, H. Xu, J. Yang, P. G. Kevrekidis, and G. Theocharis, Stability of topological edge states under strong nonlinear effects,
	    Physical Review B, 103(2), 024106 (2021).
	    \bibitem{Guo20_OL} M. Guo, S. Xia, N. Wang, D. Song, Z. Chen, and J. Yang, Weakly nonlinear topological gap solitons in Su-Schrieffer-Heeger photonic lattices,
	    Optics Letters, 45(23), 6466-6469 (2020).
	    \bibitem{Malz18_NJP} S. Malzard and H. Schomerus, Nonlinear mode competition and symmetry-protected power oscillations in topological lasers,
	    New Journal of Physics, 20(6), 063044 (2018).
	    \bibitem{Malz18_OE} S. Malzard, E. Cancellieri, and H. Schomerus, Topological dynamics and excitations in lasers and condensates with saturable gain or loss, 
	    Optics express, 26(17), 22506-22518 (2018).
	    \bibitem{Enge17_PRL} G. Engelhardt, M. Benito, G. Platero, and T. Brandes, Topologically enforced bifurcations in superconducting circuits,
	    Physical review letters, 118(19), 197702 (2017).
	    \bibitem{Kotw19_arXiv} T. Kotwal, F. Moseley, A. Stegmaier, S. Imhof, H. Brand, T. Kie{\ss}ling, R. Thomale, H. Ronellenfitsch, and J. Dunkel, Active topolectrical circuits, Proceedings of the National Academy of Sciences, 118 (32) e2106411118 (2021).

		\bibitem{syaf12} If we allow $A$ and $B$ to be complex, the 2D complex phase space is equivalent to the original 2D real phase space modulo phase rotations, so no new solution is expected.
		
		\bibitem{RoQu92} J. A. Roberts and G. R. W. Quispel, Chaos and time-reversal symmetry. Order and chaos in reversible dynamical systems,
		Physics Reports, 216(2-3), 63-177 (1992).
		
		\bibitem{LaRo98} J. S. Lamb and J. A. Roberts, Time-reversal symmetry in dynamical systems: a survey,
		Physica-Section D, 112(1), 1-39 (1998).
		
		\bibitem{C98_PhysD} A. R. Champneys, Homoclinic orbits in reversible systems and their applications in mechanics, fluids and optics,
		Physica D: Nonlinear Phenomena, 112(1-2), 158-186 (1998).
		
		\bibitem{Kevr09_Book} P. G. Kevrekidis, The discrete nonlinear Schr\"odinger equation: mathematical analysis, numerical computations and physical perspectives (Vol. 232), Springer Science \& Business Media (2009).
		
		\bibitem{bara00} I. V. Barashenkov and E. V. Zemlyanaya, Oscillatory instabilities of gap solitons: a numerical study, 
		Computer physics communications, 126(1-2), 22-27 (2000).
		
		\bibitem{derk05} G. Derks and G. A. Gottwald, A robust numerical method to study oscillatory instability of gap solitary waves,
		SIAM Journal on Applied Dynamical Systems, 4(1), 140-158 (2005).
		
		\bibitem{MaRe19_arXiv} J. L. Marzuola, M. Rechtsman, B. Osting, and M. Bandres, Bulk soliton dynamics in bosonic topological insulators,
		arXiv preprint arXiv:1904.10312 (2019).
		
		\bibitem{BiWi19_PRA} A. Bisianov, M. Wimmer, U. Peschel, and O. A. Egorov, Stability of topologically protected edge states in nonlinear fiber loops,
		Physical Review A, 100(6), 063830 (2019).
		
		\bibitem{kiri16} O. B. Kirikchi, A. A. Bachtiar, and H. Susanto, Bright Solitons in a PT-Symmetric Chain of Dimers,
		Advances in Mathematical Physics, 2016 (2016).
		
		\bibitem{LeDe16_APX} D. Leykam, and A. S. Desyatnikov, Conical intersections for light and matter waves,
		Advances in Physics: X, 1(1), 101-113 (2016).
		
		\bibitem{HaCa09_PhysD} L. H. Haddad and L. D. Carr, The nonlinear Dirac equation in Bose-Einstein condensates: Foundation and symmetries,
		Physica D: Nonlinear Phenomena, 238(15), 1413-1421 (2009).
		
		\bibitem{CuKe16_PRL} J. Cuevas-Maraver, P. G. Kevrekidis, A. Saxena, A. Comech, and R. Lan, Stability of solitary waves and vortices in a 2D nonlinear Dirac model,
		Physical Review Letters, 116(21), 214101 (2016).
		
		\bibitem{bara98} I. V. Barashenkov, D. E. Pelinovsky, and E. V. Zemlyanaya, Vibrations and oscillatory instabilities of gap solitons,
		Physical Review Letters, 80(23), 5117 (1998).
		
		\bibitem{chug06} M. Chugunova and D. Pelinovsky,  Block-diagonalization of the symmetric first-order coupled-mode system,
		SIAM Journal on Applied Dynamical Systems, 5(1), 66-83 (2006).
		
		\bibitem{Temp21_arXiv} \textcolor{black}{J. R. Tempelman, K. H. Matlack, and A. F. Vakakis, Topological Protection in a Strongly Nonlinear Interface Lattice,
		arXiv preprint arXiv:2105.08137 (2021).}
		
		\bibitem{dobr18} D. A. Dobrykh, A. V. Yulin, A. P. Slobozhanyuk, A. N. Poddubny, and Y. S. Kivshar, Nonlinear control of electromagnetic topological edge states,
		Phys. Rev. Lett. 121, 163901 (2018).
		
		\bibitem{ACZ13_PRA} M. J. Ablowitz, C. W. Curtis, and Y. Zhu, Localized nonlinear edge states in honeycomb lattices,
		Physical Review A, 88(1), 013850 (2013).
		
		\bibitem{LRPS16_PRA} Y. Lumer, M. C. Rechtsman, Y. Plotnik, and M. Segev, Instability of bosonic topological edge states in the presence of interactions,
		Physical Review A, 94(2), 021801 (2016).
		
	\end{thebibliography}
\end{document}